\documentclass[12pt]{iopart}

\usepackage{iopams}
\usepackage{graphicx}
\usepackage{subfigure}
\usepackage{mcite}

\begin{document}

\title{Absence of dynamical gap generation in suspended graphene}

\author{Jing-Rong Wang$^1$ and Guo-Zhu Liu$^{1,2,3}$\footnote[0]{$^3$Author to whom any correspondence
should be addressed.}}

\address{$^1$Department of Modern Physics, University of Science and
Technology of China, Hefei, Anhui, 230026, P.R. China}
\address{$^2$Max Planck Institut f$\ddot{u}$r Physik komplexer
Systeme, D-01187 Dresden, Germany} \ead{gzliu@ustc.edu.cn}

\begin{abstract}
There is an interesting proposal that the long-range Coulomb
interaction in suspended graphene can generate a dynamical gap,
which leads to a semimetal-insulator phase transition. We revisit
this problem by solving the self-consistent Dyson-Schwinger
equations of wave function renormalization and fermion gap. In order
to satisfy the Ward identity, a suitable vertex function is
introduced. The impacts of singular velocity renormalization and
dynamical screening on gap generation are both included in this
formalism, and prove to be very important. We obtain a critical
interaction strength, $3.2 < \alpha_{c} < 3.3$, which is larger than
the physical value $\alpha = 2.16$ for suspended graphene. It
therefore turns out that suspended graphene is a semimetal, rather
than insulator, at zero temperature.
\end{abstract}

\pacs{71.10.Hf, 71.30.+h, 73.43.Nq}

\maketitle


\section{Introduction}

After the successful fabrication of monolayer graphene in laboratory
\cite{Novoselov05}, tremendous experimental and theoretical efforts
have been devoted to exploring its novel and fascinating properties
\cite{CastroNeto, Peres, DasSarma, Kotov}. Without any interactions,
the low energy quasiparticle excitations of graphene are massless
Dirac fermions with linear dispersion \cite{Semenoff84}. The
long-range Coulomb interaction between Dirac fermions is poorly
screened due to the vanishing of density of states vanishes at Dirac
points, and is therefore anticipated to be responsible for many
unusual behaviors of graphene \cite{CastroNeto, DasSarma, Kotov}.
Interestingly, it can lead to singular renormalization of fermion
velocity \cite{Gonzalez93,Gonzalez94, Gonzalez99, Son07}, which is
able to induce unconventional properties in several observable
quantities, such as specific heat \cite{Vafek07, Sheehy},
compressibility \cite{Sheehy, HwangSh}, and electrical conductivity
\cite{HerbutMS, Fritz, Mish, Sheehy2}. It is remarkable that the
predicted singular velocity renormalization has already been
observed in recent experiments \cite{Elias}.

If the long-range Coulomb interaction is sufficiently strong, a
finite fermion gap may be dynamically generated through the
formation of excitonic particle-hole pairs
\cite{Khveshchenko01,Gorbar02, Khveshchenko04, Liu09,
Khveshchenko09, Gamayun10, Sabio, Zhang11, LiuWang, WangLiu1,
WangLiu2, Gamayun09, Wang11, Vafek08, Gonzalez10, Gonzalez11,
Drut09A, Drut09B, Drut09C, Hands08, Hands10, Giedt11, Hands11}. The
dynamical gap fundamentally changes the ground state of graphene
\cite{CastroNetoPhys}. As a consequence, there will be a quantum
phase transition from semimetal to excitonic insulator. This gap
generating mechanism is a concrete realization of the
non-perturbative phenomenon of dynamical chiral symmetry breaking
that has been extensively investigated in particle physics for five
decades \cite{Nambu, Miransky, Roberts, Appelquist88,Nash89}. From a
technological point of view, a finite gap would make graphene more
promising as a candidate material for manipulating novel electronic
devices \cite{CastroNetoPhys, Geim07}. For these reasons, the
dynamical gap generation and the resultant semimetal-insulator
transition have attracted intense theoretical interest in recent
years. Many analytical and computational tools, including
Dyson-Schwinger(DS) gap equation \cite{Khveshchenko01, Gorbar02,
Khveshchenko04, Liu09, Khveshchenko09, Gamayun10, Sabio, Zhang11,
LiuWang, WangLiu1, WangLiu2}, Bethe-Salpeter equation
\cite{Gamayun09, Wang11}, renormalization group (RG) \cite{Vafek08,
Gonzalez10, Gonzalez11}, and large scale Monte Carlo simulation
\cite{Drut09A, Drut09B, Drut09C, Hands08, Hands10, Giedt11}, have
been applied to study this problem. A critical interaction strength
$\alpha_c$ and a critical fermion flavor $N_c$ are found in almost
all these investigations: a dynamical gap is opened only when
$\alpha > \alpha_c$ and $N < N_c$. If one fixes the physical flavor
$N = 2$, the semimetal-insulator transition is turned solely by the
parameter $\alpha$, with $\alpha_c$ defining the quantum critical
point. We list some existing values of $\alpha_c$ for $N = 2$ in
Table I. With a few exceptions, $\alpha_c$ obtained in most
calculations is smaller than the physical value $\alpha = 2.16$ of
graphene suspended in vacuum. Therefore, it is suggested by many
that the suspended graphene should be an insulator at zero
temperature.

However, there is little experimental evidence for the predicted
insulating ground state. Actually, a recent experiment put an upper
limit, as small as $0.1 \mathrm{meV}$, on the possible gap in
suspended graphene \cite{Elias}. It is known that a finite gap is
observed in graphene which is placed on specific substrate
\cite{Zhou07} or has finite-size configurations
\cite{Ponomarenko08}. Nevertheless, this gap appears to be induced
by the very particular environments, and can not be regarded as a
true dynamical gap generated by chiral vacuum condensation.
Generally speaking, there may be several reasons for the lack of
clear experimental evidence for dynamical gap. For instance, the
dynamical gap can be strongly suppressed by thermal fluctuation
\cite{Khveshchenko01, Gorbar02, Khveshchenko04, Liu09}, doping
\cite{Liu09}, and/or disorders \cite{Liu09, LiuWang}, which makes it
technically hard to measure the dynamical gap in realistic
experiments. Another possibility is that the Coulomb interaction is
simply not strong enough to open a dynamical gap even in the clean
and zero-temperature limits.

\begin{table}[h]
\caption{Existing predictions for the critical interaction strength
at $N=2$. To make a comparison between the results obtained in
different references, here we define $\alpha = e^2/v_F
\varepsilon$.} \centering \setlength{\tabcolsep}{0.6pc}
\hspace{-3mm}
\begin{tabular}{llll}
\hline\hline
Reference & $\alpha_c$  & Reference & $\alpha_{c}$ \\
\hline
\cite{Khveshchenko01} \cite{Gorbar02}    & 2.33 & \cite{WangLiu2}   & 1.02\\
\cite{Khveshchenko04}                    & 1.1  & \cite{Gamayun09}  & 1.62\\
\cite{Liu09}                             & 1.2 & \cite{Vafek08}    & 0.833\\
\cite{Khveshchenko09}                    & 1.13  & \cite{Gonzalez10} & 2.5\\
\cite{Gamayun10}                         & 0.92 & \cite{Gonzalez11} & 0.99 \\
\cite{Sabio}                             & 8.1  & \cite{Drut09A}       & 1.11(6)\\
\cite{WangLiu1}                          & 1.79 & \cite{Hands10}    & 1.66\\
\hline\hline
\end{tabular}
\smallskip
\label{tab:alphac}
\end{table}

Motivated by the recent theoretical and experimental progress, we
revisit this problem in order to specify the genuine ground state of
suspended graphene. Here we only consider clean suspended graphene
at zero doping and zero temperature. The gap generation will be
examined by means of DS equation, which is known to be a very
powerful tool of analyzing dynamical gap generation in a number of
strongly interacting models \cite{Nambu, Miransky, Roberts}, such as
Nambu-Jona-Lasinio model \cite{Nambu, Miransky}, quantum
chromodynamics (QCD) \cite{Miransky, Roberts}, three-dimensional
quantum electrodynamics (QED$_3$) \cite{Appelquist88, Nash89,
Maris}, and graphene.

In Ref.~\cite{Khveshchenko01}, Khveshchenko applied the DS equation
to examine the possibility of dynamical gap generation in graphene.
In order to simplify the very complicated nonlinear gap equation, a
number of approximations are introduced. First of all, the energy
dependence of the polarization function is neglected, which is
usually called instantaneous approximation in literature
\cite{Khveshchenko01, Gorbar02, Khveshchenko04, Liu09,
Khveshchenko09, Gamayun10, Zhang11, LiuWang, WangLiu1}. In addition,
the renormalizations of wave function and fermion velocity are both
ignored. In the subsequent years, the existence of dynamical gap
generation in suspended graphene was rechecked after improving some
of these approximations \cite{Liu09, Khveshchenko09, Gamayun10,
Sabio}, and it became clear that these neglected effects can
significantly increase or decrease $\alpha_{c}$. Unfortunately, due
to the formal complexity of DS equation(s), it often happens that
one specific approximation is improved at the cost of introducing
another. In Refs.~\cite{Khveshchenko09, Sabio}, the effect of
fermion velocity renormalization on gap generation is investigated
and showed to increase $\alpha_{c}$. However, the dynamical part of
polarization function is not well addressed in their calculations.
In Ref.~\cite{Liu09}, a full dynamical polarization function is
utilized, but the velocity renormalization is not included.
Ref.~\cite{Gamayun10} performs a detailed analysis of the influence
of dynamical polarization function without including the velocity
renormalization and the energy-dependence of dynamical gap. Another
potentially important effect is the strong fermion damping caused by
Coulomb interaction. It is predicted that the long-ranged Coulomb
interaction, even though being too weak to open a gap, produces
marginal Fermi liquid behavior \cite{Gonzalez99, Khveshchenko01,
Sarma07, WangLiu10}. The strong fermion damping may affect dynamical
gap generation, but is rarely considered in the existing literature.
Since a precise value of $\alpha_{c}$ is crucial to answer the
question regarding the true ground state of graphene, it is
necessary to calculate $\alpha_{c}$ after going beyond all these
approximations.

We will show that all the aforementioned effects can be incorporated
self-consistently by constructing a set of DS equations of wave
function renormalization $A_{0,1}(p_0,\mathbf{p})$ and dynamical
fermion gap $m(p_0,\mathbf{p})$. To satisfy the Ward identity, we
also include the vertex correction to the fermion self-energy. The
dynamical fermion gap and fermion velocity renormalization can be
obtained simultaneously after solving these equations. Our numerical
calculations lead to a critical coupling $3.2 < \alpha_c < 3.3$ for
physical flavor $N = 2$. Apparently, $\alpha_c$ is larger than
$\alpha = 2.16$, so the ground state of suspended graphene seems to
be a semimetal, rather than an insulator. Our $\alpha_c$ is quite
different from those obtained in previous DS equation studies, which
implies that an appropriate treatment of velocity renormalization
and dynamical screening is very important. Moreover, from
$A_{0,1}(p_0,\mathbf{p})$, we recover the singular renormalization
of fermion velocity.

The rest of the paper is organized as follows. In Sec.2, we
construct and numerically solve the self-consistent DS equations. A
critical interaction strength $\alpha_c$ is obtained from the
solutions. In Sec.3, we address the feedback effect of fermion
velocity renormalization on the effective Coulomb interaction. We
briefly summarize our results and discuss the implications in Sec.4.

\section{Dyson-Schwinger equations for gap and wave functions}

The Hamiltonian for interacting Dirac fermions is
\begin{eqnarray}
H &=& v_{F}\sum_{i=1}^{N}\int_{\mathbf{r}}
\bar{\psi}_{i}(\mathbf{r})i\mathbf{\gamma}\cdot\mathbf{\nabla}
\psi_{i}(\mathbf{r})  + \frac{v_{F}}{4\pi}\sum_{i,j}^{N}
\int_{\mathbf{r},\mathbf{r}^{\prime}} \rho_{i}(\mathbf{r})
\frac{g}{|\mathbf{r}-\mathbf{r}^{\prime}|}\rho_{j}(\mathbf{r}'),
\end{eqnarray}
where the density operator $\rho_{i}(\mathbf{r}) =
\bar{\psi}_{i}(\mathbf{r})\gamma_{0}\psi_{i}(\mathbf{r})$ and $g =
2\pi e^2/v_{F}\varepsilon$. The Dirac fermions have totally eight
indices: two sublattices, two spins, two valleys. As usual, we adopt
a four-component spinor field $\psi$ to describe the Dirac fermions
\cite{Khveshchenko01, Gorbar02}, and define its conjugate field as
$\bar{\psi} = \psi^{\dagger} \gamma_{0}$. Now the physical flavor of
fermions is $N = 2$. The $4\times 4$ $\gamma$-matrices,
$\gamma_{0,1,2}$, satisfy the Clifford algebra \cite{Khveshchenko01,
Gorbar02}. It is easy to check that the total Hamiltonian possesses
a continuous chiral symmetry, $\psi \rightarrow
e^{i\theta\gamma_5}\psi$, where the matrix $\gamma_5$ anticommutes
with $\gamma_{0,1,2}$. If the massless Dirac fermions acquire a
finite mass gap due to excitonic condensation, $\langle
\bar{\psi}\psi\rangle \neq 0$, then this continuous chiral symmetry
will be dynamically broken.

It is convenient to characterize the interaction strength by an
effective fine structure constant, $\alpha = e^2/v_F \varepsilon$.
We need to consider the strong-coupling regime since the dynamical
gap can only be opened by strong interaction. When $\alpha$ is close
to or larger than unity, the conventional perturbation expansion in
terms of $\alpha$ breaks down and one has to use $1/N$ as the
expansion parameter.

\begin{figure}[htbp]
\center
\includegraphics[width=3.2in]{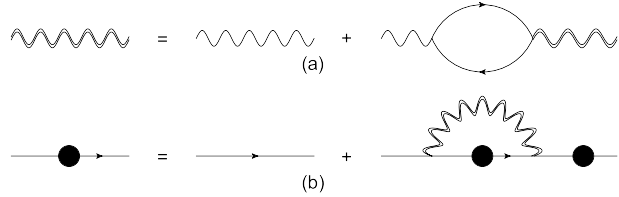}
\caption{(a) The DS equation for effective Coulomb interaction. (b)
The DS equation for fermion propagator.} \label{fig:PolaSelfEnergy}
\end{figure}

The free propagator for massless Dirac fermion is
\begin{eqnarray}
G_{0}(p_{0},\mathbf{k}) = \frac{1}{i\gamma_{0} p_{0} -
v_{F}\mathbf{\gamma}\cdot\mathbf{p}} \label{eqn:FreeFermion}.
\end{eqnarray}
Due to the Coulomb interaction, this propagator will receive
self-energy corrections and become
\begin{eqnarray}
G(p_{0},\mathbf{p}) &=& \frac{1}{i A_{0}(p_{0},\mathbf{p})\gamma_{0}
p_{0} - v_{F}A_{1}(p_{0},\mathbf{p})\mathbf{\gamma} \cdot
\mathbf{p}-m(p_{0},\mathbf{p})} \label{eqn:DressedFermion},
\end{eqnarray}
where $A_{0,1}(p_{0},\mathbf{p})$ are the wave function
renormalization and $m(p_{0},\mathbf{p})$ is the fermion gap
function. Here we would like to make a comparison with the gap
generation problem of QED$_3$. In QED$_3$, the explicit Lorentz
invariance ensures that $A_{0}(p_{0},\mathbf{p}) =
A_{1}(p_{0},\mathbf{p})$ \cite{Appelquist88,Nash89}. On the
contrary, the Lorentz invariance is explicitly broken in the present
system, so $A_{0}(p_{0},\mathbf{p}) \neq A_{1}(p_{0},\mathbf{p})$
and the fermion velocity is renormalized. The renormalized velocity
is given by $A_{1}(p_{0},\mathbf{p})/A_{0}(p_{0},\mathbf{p})$. The
function $A_{0}(p_{0},\mathbf{p})$ itself is also important since it
determines the fermion damping rate caused by the Coulomb
interaction. The dynamical fermion gap is described by
$m(p_{0},\mathbf{p})$, which acquires a finite value when excitonic
particle-hole bound states are formed due to sufficiently strong
interaction. In previous DS equation analysis, a common
approximation is to simply assume that, $A_{0}(p_{0},\mathbf{p}) =
A_{1}(p_{0},\mathbf{p}) = 1$, which drastically simplify the
complicated DS equations. However, the effects of velocity
renormalization and fermion damping can not be well addressed by
doing so. In order to include these effects in a self-consistent
manner, it is better to analyze the coupled equations of
$A_{0}(p_{0},\mathbf{p})$, $A_{1}(p_{0},\mathbf{p})$, and
$m(p_{0},\mathbf{p})$.

According to the Feynman diagram shown in Fig.1(b), the relationship
between the free and renormalized fermion propagators is formally
determined by the following DS equation,
\begin{eqnarray}
\fl G^{-1}(p_{0},\mathbf{p}) &=& G_{0}^{-1}(p_{0},\mathbf{p})+\int
\frac{dk_{0}}{2\pi}\frac{d^{2}\mathbf{k}}{(2\pi)^{2}}
\Gamma_{0}(p_{0},\mathbf{p};k_{0},\mathbf{k})
G(k_{0},\mathbf{k})\gamma_{0} V(p_{0} -
k_{0},\mathbf{p}-\mathbf{k}),\label{eqn:Dyson}
\end{eqnarray}
where $\Gamma_{0}(p_{0},\mathbf{p};k_{0},\mathbf{k})$ is the full
vertex function and $V(p_{0}-k_{0},\mathbf{p}-\mathbf{k})$ is the
effective Coulomb interaction function. Regarding the Coulomb
potential as the time component of a gauge potential, one can obtain
a Ward identity \cite{Gonzalez99} that connects the fermion
propagator and the vertex function,
\begin{eqnarray}
\Gamma_{0}\left(p_{0},\mathbf{p};p_{0},\mathbf{p}\right) =
\frac{\partial G^{-1}\left(p_{0},\mathbf{p}\right)}{i\partial
p_{0}}.
\end{eqnarray}
Apparently, $\Gamma_{0} = \gamma_{0}$ only when
$A_{0}(p_{0},\mathbf{p}) = 1$. As we are willing to examine the
effects of wave function renormalization on dynamical gap
generation, the vertex function $\Gamma_0$ can not be simply
replaced by the bare vertex $\gamma_0$. In principle, one could
build an equation of
$\Gamma_{0}\left(p_{0},\mathbf{p};k_{0},\mathbf{k}\right)$ and
couple it self-consistently to Eq.(4). However, this will make the
problem intractable. A more practical way is to assume a proper
\emph{Ansatz}. Choosing a suitable vertex function is a highly
nontrivial problem, and have been investigated extensively in the
contexts of various quantum field theories \cite{Roberts}. The
research experience that has been accumulated in QED$_3$
\cite{Maris} is especially helpful since this model is very similar
in structure to our present model. There are several frequently used
\emph{Ansatze} for the vertex function \cite{Maris}. Among these
\emph{Ansatze}, for computational convenience we choose the
following one,
\begin{eqnarray}
\Gamma_{0}\left(p_{0},\mathbf{p};k_{0},\mathbf{k}\right) &=&
f\left(p_{0},\mathbf{p};k_{0},\mathbf{k}\right)\gamma_0 =
\frac{1}{2}\left[A_{0}(p_{0},\mathbf{p}) +
A_{0}(k_{0},\mathbf{k})\right]\gamma_{0}.
\end{eqnarray}
Other possible \emph{Ansatze} of
$\Gamma_{0}\left(p_{0},\mathbf{p};k_{0},\mathbf{k}\right)$ can be
analyzed by the same procedure.

Substituting the full fermion propagator $G(p_{0},\mathbf{p})$ into
the above DS equation and then taking trace on both sides, we can
obtain the equation for dynamical gap $m(p_{0},\mathbf{p})$. In
order to derive the equations of $A_{0}(p_{0},\mathbf{p})$ and
$A_{1}(p_{0},\mathbf{p})$, we multiply both sides of
Eq.(\ref{eqn:Dyson}) by $\gamma_0 p_0$ and $\mathbf{\gamma}\cdot
\mathbf{p}$ respectively, and then take trace on both sides. After
these manipulations, we finally arrive in a set of self-consistently
coupled integral equations,

\begin{eqnarray}
\fl A_{0}(p_{0},\mathbf{p})=1+\frac{1}{2p_{0}}\int_{0}^{+\infty}
\frac{dk_{0}}{2\pi}\int\frac{d^{2}\mathbf{k}}{(2\pi)^{2}}\frac{
\left[A_{0}(p_{0},\mathbf{p}) +
A_{0}(k_{0},\mathbf{k})\right]A_{0}(k_{0},\mathbf{k})k_{0}} {
A_{0}^{2}(k_{0},\mathbf{k})k_{0}^{2} +
v_{F}^{2}A_{1}^{2}(k_{0},\mathbf{k})|\mathbf{k}|^{2} +
m^{2}(k_{0},\mathbf{k})}\nonumber
\\
\times\left[V(p_{0}+k_{0},\mathbf{p}-\mathbf{k}) -
V(p_{0}-k_{0},\mathbf{p}-\mathbf{k})\right],
\\
\fl  A_{1}(p_{0},\mathbf{p}) =
1+\frac{1}{2|\mathbf{p}|^{2}}\int_{0}^{+\infty}
\frac{dk_{0}}{2\pi}\int\frac{d^{2}\mathbf{k}}{(2\pi)^{2}}
\frac{\left[A_{0}(p_{0},\mathbf{p}) +
A_{0}(k_{0},\mathbf{k})\right]A_{1}(k_{0},\mathbf{k})
\left(\mathbf{p}\cdot\mathbf{k}\right)}
{A_{0}^{2}(k_{0},\mathbf{k})k_{0}^{2} +
v_{F}^{2}A_{1}^{2}(k_{0},\mathbf{k})|\mathbf{k}|^{2} +
m^{2}(k_{0},\mathbf{k})}\nonumber
\\
\times\left[V(p_{0}+k_{0},\mathbf{p}-\mathbf{k}) +
V(p_{0}-k_{0},\mathbf{p}-\mathbf{k})\right],
\\
\fl  m(p_{0},\mathbf{p}) =\frac{1}{2}\int_{0}^{+\infty}
\frac{dk_{0}}{2\pi}\int\frac{d^{2}\mathbf{k}}{(2\pi)^{2}}
\frac{\left[A_{0}(p_{0},\mathbf{p}) +
A_{0}(k_{0},\mathbf{k})\right]m(k_{0},\mathbf{k})}{A_{0}^{2}(k_{0},\mathbf{k})
k_{0}^{2} + v_{F}^{2}A_{1}^{2}(k_{0},\mathbf{k})|\mathbf{k}|^{2} +
m^{2}(k_{0},\mathbf{k})} \nonumber \\
\times\left[V(p_{0}+k_{0},\mathbf{p}-\mathbf{k}) +
V(p_{0}-k_{0},\mathbf{p}-\mathbf{k})\right].
\end{eqnarray}

During the derivation of these equations, we have used the following
symmetries
\begin{eqnarray}
A_{0,1}(p_{0},\mathbf{p}) &=& A_{0,1}(-p_{0},\mathbf{p}),
\\
m(p_{0},\mathbf{p})&=&m(-p_{0},\mathbf{p}),
\end{eqnarray}
which originate from the particle-hole symmetry of undoped graphene.

We now consider the effective interaction function
$V(p_{0}-k_{0},\mathbf{p}-\mathbf{k})$. The bare Coulomb interaction
is
\begin{equation}
V_{0}(\mathbf{q}) = \frac{2\pi\alpha v_{F}}{|\mathbf{q}|},
\end{equation}
with effective fine structure constant,
$\alpha=\frac{e^{2}}{v_{F}\epsilon}$. The value of dielectric
constant $\epsilon$ depends on the substrate on which the graphene
is placed. For graphene suspended in vacuum, the corresponding value
is roughly $\alpha = 2.16$; for graphene placed on SiO$_{2}$
substrate, $\alpha = 0.79$ \cite{CastroNetoPhys}. Besides the static
screening due to substrate, the Coulomb interaction is also
dynamically screened by the collective particle-hole excitations.
After including the dynamical screening, as displayed in
Fig.(\ref{fig:PolaSelfEnergy}a), the effective interaction function
becomes
\begin{equation}
V(q_0, \mathbf{q}) = \frac{1}{V_{0}^{-1}(\mathbf{q}) +
\Pi(q_{0},\mathbf{q})}.\label{eqn:DressedCoulomb},
\end{equation}
The dynamical polarization function $\Pi(q)$ is defined as
\begin{eqnarray}
\Pi(q_{0},\mathbf{q}) = -N \int\frac{d^{3}k}{(2\pi)^{3}}\mathrm{Tr}
\left[\gamma_{0}G_0(k_{0},\mathbf{q})\gamma_{0}G_0(k_{0}+q_{0},\mathbf{k}+\mathbf{q})\right]
\end{eqnarray}
to the leading order, which amounts to random phase approximation
(RPA). It is straightforward to obtain
\begin{equation}
\Pi(q_{0},\mathbf{q}) = \frac{N}{8}\frac{\left|\mathbf{q}\right|^2}
{\sqrt{q_{0}^{2}+v_{F}^{2}|\mathbf{q}|^2}}.
\end{equation}
Apparently, although the bare Coulomb interaction
$V_{0}(\mathbf{q})$ is independent of energy, the effective
interaction function $V(q_0, \mathbf{q})$ depend on both energy
$q_0$ and momentum $\mathbf{q}$ due to dynamical screening. This
$V(q_0, \mathbf{q})$ should be substituted to the self-consistent
equations (7-9).

Before doing numerical calculations, it is interesting to compare
the present problem with QED$_3$. In QED$_3$, the inverse of bare
photon propagator is $\propto q^2$, with $q$ being three-dimensional
energy-momentum. The corresponding polarization is known to be
$\propto |q|$, which is larger than $q^2$ at low energies
\cite{Appelquist88}. Therefore, the effective photon propagator is
dominated by the polarization and the bare term is relatively
unimportant. In the present problem, the inverse of bare Coulomb
interaction is $V_{0}^{-1}(\mathbf{q}) \propto |\mathbf{q}|$. The
polarization $\Pi(q_{0},\mathbf{q})$ is more complex than that in
QED$_3$, but is evidently linear in momentum $|\mathbf{q}|$.
Therefore, the bare term $V_{0}^{-1}(\mathbf{q})$ and the
polarization $\Pi(q_{0},\mathbf{q})$ appearing in the denominator of
effective interaction $V(q_0, \mathbf{q})$ are equally important,
which is different from QED$_3$.

\begin{figure}[htbp]
\center
\includegraphics[width=3.3in]{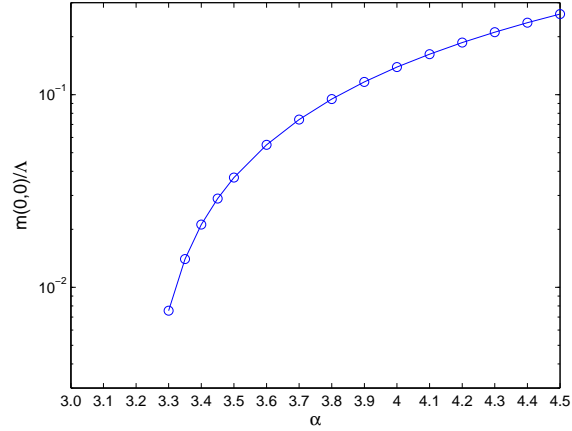}
\caption{Relationship between dynamical gap $m(0,0)$ and $\alpha$.
Apparently, $m(0,0)$ decreases as $\alpha$ decreases. $m(0,0)$
becomes zero once $\alpha$ decreases below certain critical value
$\alpha_c$. It is technically hard to get a precise $\alpha_c$.
Numerical calculations show that $m(0,0)$ remains finite at $\alpha
= 3.3$ but vanishes at $\alpha = 3.2$, so the critical value must be
$3.2 < \alpha_c < 3.3$.} \label{fig:Gap0}
\end{figure}

\begin{figure*}[htbp]
\center
   \includegraphics[width=5.5in]{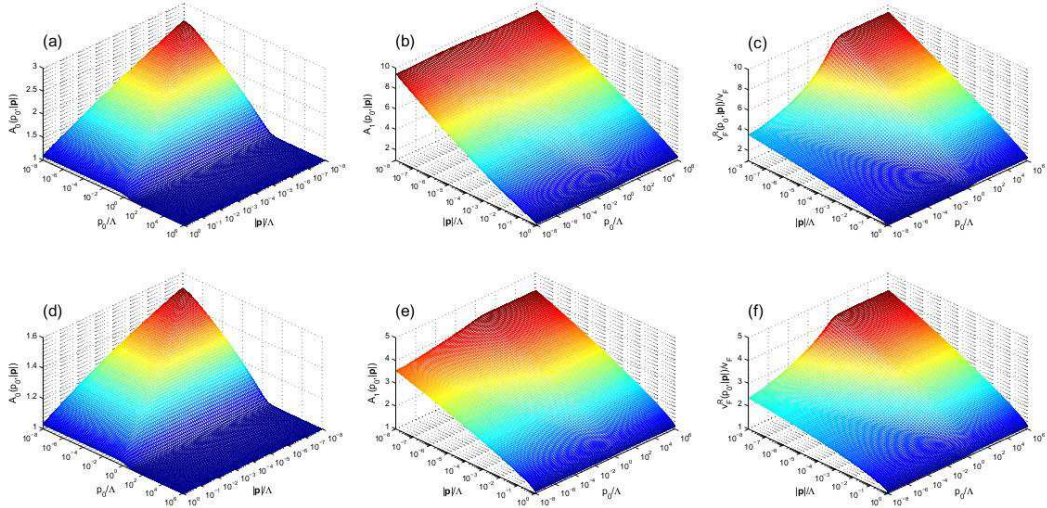}
\caption{(a), (b), (c) are $A_{0}(p_{0},|\mathbf{\mathbf{p}}|)$,
$A_{1}(p_{0},|\mathbf{p}|)$, $v_{F}^{R}(p_{0},|\mathbf{p}|)$ for the
graphene suspended in vacuum with $\alpha=2.16$. (d), (e), (f) are
$A_{0}(p_{0},|\mathbf{\mathbf{p}}|)$, $A_{1}(p_{0},|\mathbf{p}|)$,
$v_{F}^{R}(p_{0},|\mathbf{p}|)$ for the graphene placed on SiO$_{2}$
substrate with $\alpha=0.79$.} \label{fig:ReFun}
\end{figure*}

After time-consuming but straightforward numerical calculations, we
obtain the relationship between $m(0,0)$ and $\alpha$, shown in
Fig.(\ref{fig:Gap0}). It is difficult to get a precise $\alpha_c$
because the necessary computational time becomes increasingly long
as $\alpha$ is approaching its critical value. Therefore, here we
can only give a narrow range of $\alpha_c$. For physical fermion
flavor $N = 2$, the critical strength is $3.2 < \alpha_c < 3.3$.
This critical value is obviously larger than $\alpha = 2.16$, so it
turns out that the zero-temperature ground state of suspended
graphene is a semimetal in the clean limit. This result is
consistent with the fact that so far no experimental evidence for
insulating ground state of suspended graphene has been reported.

Our critical value $\alpha_c$ is very different from the values
presented in earlier publications. Compared with the previous gap
equation computations, the key improvements in our analysis are that
we have included the wave function renormalization
$A_{0,1}(p_{0},\mathbf{p})$ self-consistently and maintained the
energy-dependence of all the functions of
$A_{0,1}(p_{0},\mathbf{p})$, $m(p_{0},\mathbf{p})$, and
$\Pi(q_{0},\mathbf{q})$. We also have adopted an \emph{Ansatz} for
the vertex function in order not to violate the Ward identity. Our
results indicate that fermion velocity renormalization, fermion
damping, and dynamical screening of Coulomb interaction all play
important roles in determining the critical interaction strength
$\alpha_c$.

It is now useful to make a more concrete comparison between our
$\alpha_c$ and those presented in some recent literature. For
example, our $\alpha_c$ is much larger than the $\alpha_c$ obtained
by Liu \emph{et al.} \cite{Liu09} and Gamayun \emph{et al.}
\cite{Gamayun10}, who considered a fully dynamical polarization but
ignored wave functions and velocity renormalization. Sabio \emph{et.
al.} \cite{Sabio} studied dynamical gap generation by means of a
variational method that naturally includes velocity renormalization,
and found a critical value $\alpha_c = 8.1$, which is much greater
than our $\alpha_c$. Such a large quantitative difference is
presumably owing to the instantaneous approximation of Coulomb
interaction adopted in their analysis. Once the instantaneous
approximation is assumed, the effective Coulomb interaction no
longer depends on energy, and $V(p_{0}+k_{0},\mathbf{p}-\mathbf{k})$
will be identically equal to $V(p_{0}-k_{0},\mathbf{p}-\mathbf{k})$.
As shown in Eq.(7), the time component of wave function becomes
$A_0(p_0,\mathbf{p}) = 1$. Now the fermion damping effect is
automatically neglected and the velocity renormalization is solely
determined by $A_1(p_0,\mathbf{p})$. Moreover, $A_0(p_0,\mathbf{p})
= 1$ implies that vertex function is simply $\Gamma_0 = \gamma_0$.
After these simplifications, our DS equations become similar to
those presented in Ref.~\cite{Sabio}. By solving these simplified DS
equations, we find no evidence for dynamical gap even for infinite
coupling $\alpha \rightarrow \infty$. This indicates that the
instantaneous approximation misses very important fluctuation
effects. However, despite the large difference in the magnitude of
$\alpha_c$, our conclusion do agree with Sabio \emph{et al.} that no
dynamical gap is generated in suspended graphene.

\begin{figure}[htbp]
\center
    \subfigure{
   \includegraphics[width=3in]{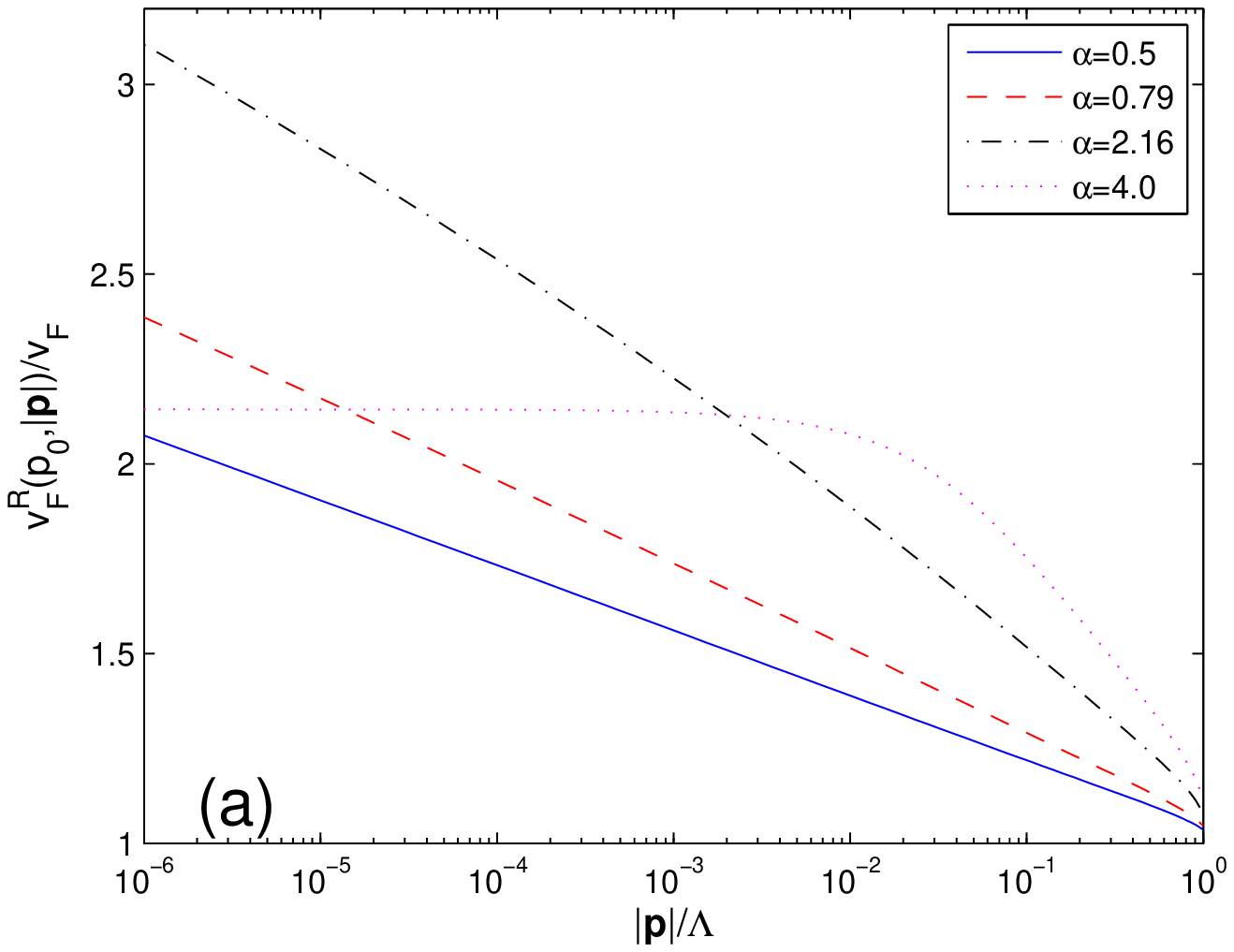}}
   \subfigure{
   \includegraphics[width=3in]{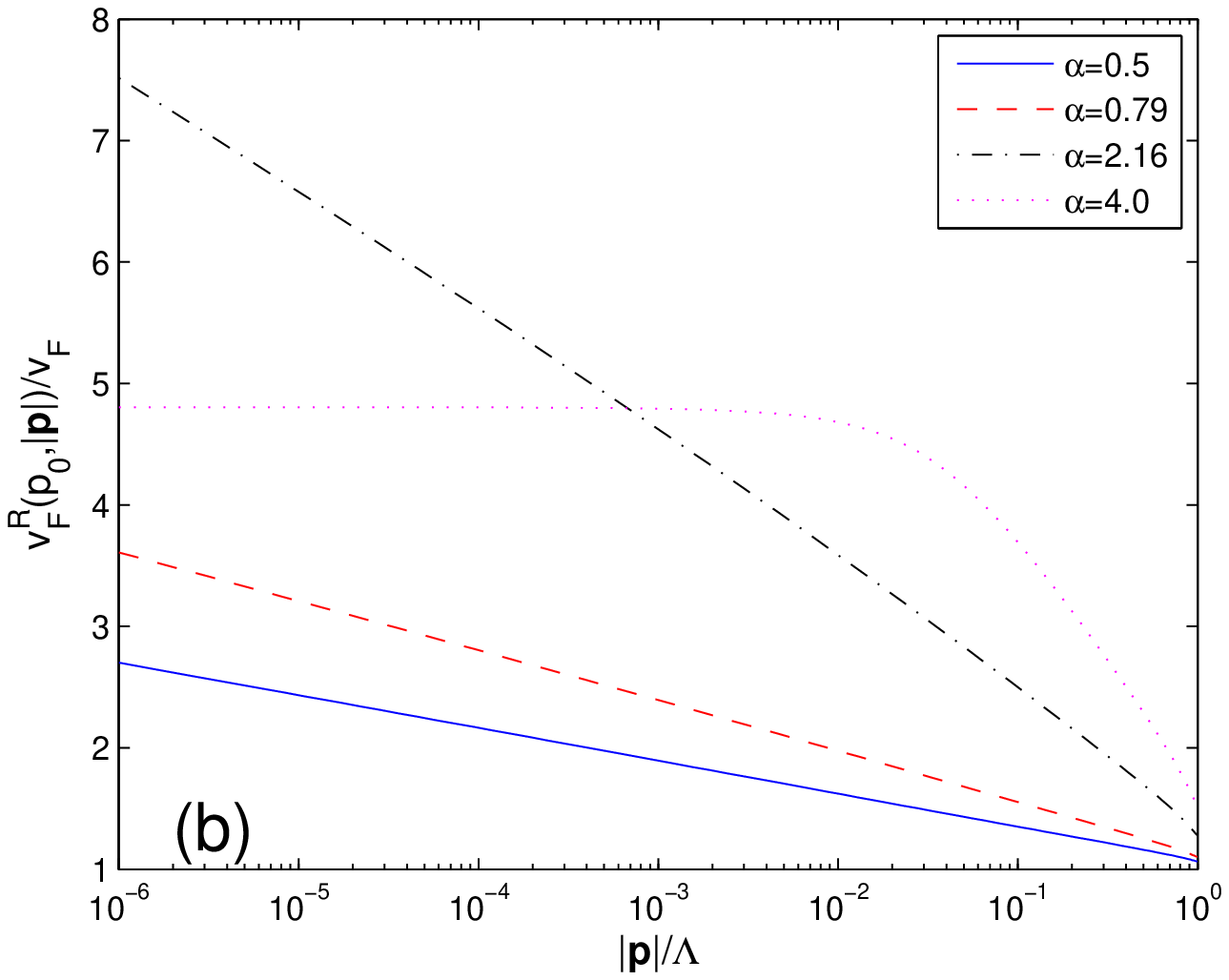}}
\caption{(a) Renormalized velocity $v_{F}^{R}(p_{0},\mathbf{p})$ in
the $p_{0}\rightarrow 0$ limit for different values of $\alpha$; (b)
Renormalized velocity $v_{F}^{R}(p_{0},\mathbf{p})$ in the
$p_{0}\rightarrow \infty$ limit for different values of $\alpha$.}
\label{fig:VFLH}
\end{figure}

Although the Coulomb interaction in suspended graphene is too weak
to generate a dynamical gap, it still leads to unusual properties:
strong fermion damping and singular velocity renormalization.
Usually, the fermion damping rate and the fermion velocity
renormalization are calculated separately. Our DS equation approach
includes the mutual influence of these two quantities
self-consistently, so the damping rate and velocity renormalization
can be simultaneously obtained from the solutions of
$A_{0}(p_{0},\mathbf{p})$ and $A_{1}(p_{0},\mathbf{p})$. The energy
and momentum dependence of $A_{0}(p_{0},\mathbf{p})$ and
$A_{1}(p_{0},\mathbf{p})$ are shown in Fig.(\ref{fig:ReFun}), with
(a-b) for suspended graphene with $\alpha = 2.16$ and (d-e) for
graphene placed on SiO$_{2}$ substrate with $\alpha = 0.79$. It is
clear that both $A_{0}(p_{0},\mathbf{p})$ and
$A_{1}(p_{0},\mathbf{p})$ increase as interaction strength $\alpha$
is increasing. $A_{0}(p_{0},\mathbf{p})$ decreases with growing
energy and momentum. Different from $A_0(p_{0},\mathbf{p})$,
$A_{1}(p_{0},\mathbf{p})$ is a decreasing function of momentum
$|\mathbf{p}|$ but an increasing function of energy $p_{0}$. When
$|\mathbf{p}|$ decreases gradually, $A_{1}(p_{0},\mathbf{p})$ keeps
increasing monotonically; when $p_{0} \rightarrow 0$ or $p_{0}
\rightarrow \infty$, $A_{1}(p_{0},|\mathbf{p}|)$ saturates to a
finite value $A_{1}(0,\mathbf{p})$ or $A_{1}(\infty,\mathbf{p})$,
respectively.

The fermion damping is determined by the time component of wave
function renormalization, $A_{0}(p_{0},\mathbf{p})$. For any given
energy $p_{0}$, $A_{0}(p_{0},\mathbf{p})$ increases as momentum
$|\mathbf{p}|$ decreases in the region $|\mathbf{p}|
> p_{0}$, but saturates to a finite value
$A_{0}(p_{0},0)$ as $|\mathbf{p}|$ decreases in the region
$|\mathbf{p}| < p_{0}$. For given momentum $|\mathbf{p}|$, when
$p_{0}$ decreases gradually, $A_{0}(p_{0},\mathbf{p})$ increases
with decreasing $p_0$ in the region $p_{0} > |\mathbf{p}|$, but it
saturates to a finite value $A_{0}(0,\mathbf{p})$ with decreasing
$p_0$ in the region $p_{0} < |\mathbf{p}|$. In the low energy
regime, $A_{0}(p_{0},\mathbf{p})$ can be approximated as
$A_{0}(p_{0},\mathbf{p}) \sim
1+F(\Lambda/\sqrt{p_{0}^{2}+\mathbf{p}^2})$, where $F(x)$ is an
increasing function of $x$. In principle, the fermion damping rate
can be obtained from $A_{0}(p_{0},\mathbf{p})$ by making analytic
continuation, $ip_{0} \rightarrow \omega+i\delta$. Unfortunately,
the function $F(\Lambda/\sqrt{p_{0}^2+\mathbf{p}^2})$ is formally
quite complicated and can not be written in terms of a simple
analytical formula. Our numerical results suggest that
$F(\Lambda/\sqrt{p_{0}^2+\mathbf{p}^2})$ grows a little more slowly
than $\log(\Lambda/\sqrt{p_{0}^2 + \mathbf{p}^2})$ as $\sqrt{p_{0}^2
+ \mathbf{p}^2}$ decreases continuously, hence the damping rate does
not display an exact linear dependence on energy. By ignoring the
singular velocity renormalization, some pervious perturbative
\cite{Gonzalez99, Sarma07} and self-consistent \cite{Khveshchenko01,
WangLiu10} calculations predict a marginal Fermi liquid behavior,
namely $\Gamma(\omega) \sim \omega$. Such linear-in-energy behavior
disappears once the singular velocity renormalization is
self-consistently considered. However, the renormalization factor
$Z$ extracted from our numerical results is found to vanish, $Z =
0$, so the suspended graphene does not have well-defined
quasiparticles. In this sense, our results are qualitatively
consistent with the previous conclusions \cite{Khveshchenko01,
Sarma07, WangLiu10}.

\begin{figure*}[htbp]
\center
   \includegraphics[width=5in]{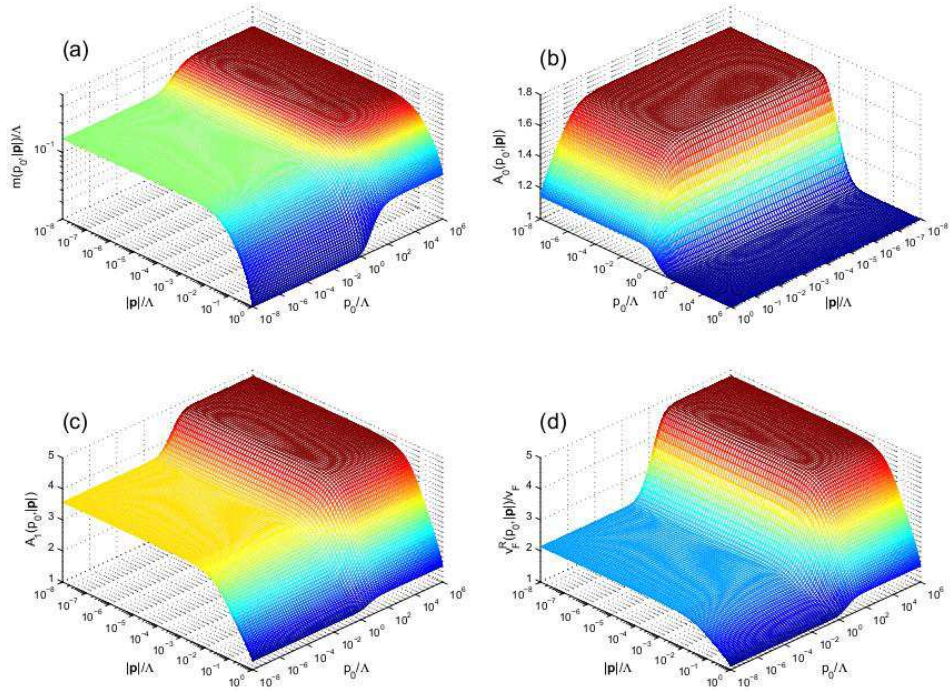}
\caption{(a), (b), (c) and (d) are
$m(p_{0},|\mathbf{p}|)$,$A_{0}(p_{0},|\mathbf{p}|)$,
$A_{1}(p_{0},|\mathbf{p}|)$ and $v_{F}^{R}(p_{0},|\mathbf{p}|)$ for
the graphene with $\alpha=4.0$.} \label{fig:Gap}
\end{figure*}

The fermion velocity renormalization is related to many properties
of graphene \cite{Kotov, Gonzalez93,Gonzalez94, Son07, Vafek07,
Sheehy, HwangSh, HerbutMS, Fritz, Mish, Sheehy2}. Such effect is
usually studied by RG method \cite{Gonzalez93, Gonzalez94,
Gonzalez99, Son07}. It can be naturally obtained by solving the
self-consistent DS equations. The ratio between renormalized and
bare fermion velocities is defined as
\begin{eqnarray}
\frac{v_{F}^{R}(p_{0},\mathbf{p})}{v_{F}} =
\frac{A_{1}(p_{0},\mathbf{p})}{A_{0}(p_{0},\mathbf{p})}.
\end{eqnarray}
We plot $v_{F}^{R}(p_{0},\mathbf{p})$ for $\alpha = 2.16$ and
$\alpha = 0.79$ in (c) and (f) of Fig.(\ref{fig:ReFun}),
respectively. The renormalized fermion velocity
$v_{F}^{R}(p_{0},\mathbf{p})$ increases singularly as $|\mathbf{p}|
\rightarrow 0$, which reproduces the previous results of RG analysis
\cite{Gonzalez93, Gonzalez94, Gonzalez99, Son07}. We also obtain the
energy dependence of renormalzied velocity
$v_{F}^{R}(p_{0},\mathbf{p})$, which is not well captured in RG
analysis. For given momentum, $v_{F}^{R}(p_{0},\mathbf{p})$ is an
increasing function of energy. When $p_{0} \rightarrow 0$ or $p_{0}
\rightarrow \infty$, $v_{F}^{R}(p_{0},\mathbf{p})$ ceases to change,
but is saturated to $v_{F}^{R}(0,\mathbf{p})$ or
$v_{F}^{R}(\infty,\mathbf{p})$, as can be seen from Fig.(3c) and
Fig.(3f) as well as Fig.(4).

It is also interesting to consider the gapped phase with $\alpha >
\alpha_c$, though no known graphene material manifests such a large
$\alpha$. When $\alpha = 4.0$, a dynamical gap is generated by the
Coulomb interaction. The corresponding $m(p_{0},\mathbf{p})$,
$A_{0,1}(p_{0},\mathbf{p})$, and $v_{F}^{R}(p_{0},\mathbf{q})$ are
shown in Fig.(\ref{fig:Gap}). The gap $m(p_{0},\mathbf{p})$
decreases with growing momentum but increases with growing energy.
These behaviors are very different from those in QED$_{3}$ where the
dynamical gap is a decreasing function of both energy and momentum.
This difference reflects the different energy-dependence of Coulomb
interaction and gauge interaction. From
Eq.(\ref{eqn:DressedCoulomb}), we know that the effective Coulomb
interaction decreases with growing momentum but increases with
growing energy. However, the gauge interaction in QED$_{3}$ always
decreases as energy or momentum grows, which is in accordance with
the fact that QED$_{3}$ exhibits asymptotic freedom
\cite{Appelquist88,Nash89}.

In the low-energy regime, $\sqrt{p_{0}^{2}+|\mathbf{p}|^{2}} <
m(0,0)$, $A(p_{0},\mathbf{p})$ saturates to a finite value and can
be approximate as $A_{0}(p_{0},\mathbf{p})\sim 1 +
F(\Lambda/\sqrt{p_{0}^{2}+|\mathbf{p}|^{2}+m^2(0,0)})$ with $F(x)$
being certain increasing function of $x$. After analytic
continuation, $ip_{0} \rightarrow \omega + i\delta$, the fermion
damping rate is found to vanish when $|\omega| <
\sqrt{|\mathbf{p}|^2 + m^2(0,0)}$. It implies that the fermions can
not be excited below the scale of $m(0,0)$. Unlike the case of
semimetal phase, $v_{F}^{R}(p_{0},\mathbf{p})$ does not keep
increasing as momentum decreases. For any given energy,
$v_{F}^{R}(p_{0},\mathbf{p})$ saturates to $v_{F}^{R}(p_{0},0)$
below the scale $|\mathbf{p}|\sim m(0,0)$, so the singular velocity
renormalization is suppressed by the dynamical gap generation in the
insulating phase, as shown in Fig.(4). These results are consistent
with those obtained by perturbative calculations \cite{Kotov08} and
by functional RG calculations \cite{Sinner10}.

\section{Effects of velocity renormalization on polarization}

In the calculations presented above, we have used the RPA expression
of the dynamical polarization function $\Pi(q_{0},\mathbf{q})$, in
which the fermion velocity is a constant. However, this function may
receive sizeable feedback effects from the velocity renormalization.
Since the effective Coulomb interaction plays a crucial role in
determining $\alpha_c$, it deserves to examine these effects. In a
fully self-consistent analysis, one should consider the following
defination
\begin{eqnarray}
\Pi(q_{0},\mathbf{q}) = - N\int\frac{d^{3}k}{(2\pi)^{3}}\mathrm{Tr}
\left[\Gamma_{0}G(k_{0},\mathbf{k})\gamma_{0}G(k_{0}+q_{0},\mathbf{k}+\mathbf{q})\right],
\end{eqnarray}
where $G(k_{0},\mathbf{k})$ is the full fermion propagator given by
Eq.(3) and $\Gamma_0$ is the vertex function. By doing so, the
polarization $\Pi(q_{0},\mathbf{q})$ couples self-consistently to
the equations of $A_{0,1}(p_{0},\mathbf{p})$ and
$m(p_{0},\mathbf{p})$. Unfortunately, this would make the
computational time unacceptably long since the integrations over
energy and momenta must be performed separately in the present
problem. To simplify the calculations, we assume that the RPA
expression of $\Pi(q_{0},\mathbf{q})$ provides a reliable
description of the dynamical screening effect, which then allows us
to replace the constant velocity $v_F$ appearing in
$\Pi(q_{0},\mathbf{q})$ by the renormalized velocity
$v_{F}^{R}(p_{0},\mathbf{p})$. This strategy is analogous to that
employed in Ref.~\cite{Khveshchenko09}. After this substitution, the
effective interaction becomes
\begin{equation}
V(q_0,\mathbf{q}) = \frac{1}{\frac{|\mathbf{q}|}{2\pi \alpha v_F} +
\frac{N}{8}\frac{\left|\mathbf{q}\right|^2} {\sqrt{q_{0}^{2} +
(v_{F}^{R}(q_{0},\mathbf{q}))^2|\mathbf{q}|^2}}}.
\end{equation}
It is larger than
\begin{equation}
\frac{1}{\frac{|\mathbf{q}|}{2\pi \alpha v_F} +
\frac{N}{8}\frac{\left|\mathbf{q}\right|^2} {\sqrt{q_{0}^{2} +
v_F^{2}|\mathbf{q}|^2}}}
\end{equation}
since $v_{F}^{R}(q_{0},\mathbf{q}) > v_F$, so the velocity
renormalization tends to enhance the effective interaction strength.
In order to examine the extend to which this feedback effect changes
$\alpha_c$, we re-solve the coupled gap equations after substituting
Eq.(18) into Eqs.(7 - 9). We find the fermion gap is always zero
when $\alpha \leq 2.9$. It implies that such feedback effect does
not change $\alpha_c$ substantially. In particular, $\alpha_c$
obtained after considering this effect is still larger than the
physical value $\alpha = 2.16$ in suspended graphene.

\section{Summary and discussions}

In conclusion, we have presented a detailed analysis of dynamic gap
generation due to Coulomb interaction in graphene using the coupled
DS equations of wave function renormalization and fermion gap. After
including the effects of fermion velocity renormalization and
dynamical screening of Coulomb interaction, we obtain a critical
interaction strength, $3.2 < \alpha_c < 3.3$. It is apparently
greater than the physical strength $\alpha = 2.16$ in suspended
graphene. Therefore, the Coulomb interaction in suspended graphene
is too weak to generate a dynamical fermion gap, and the semimetal
ground state of suspended graphene is robust, which is consistent
with the fact that so far no insulating phase has been observed in
experiments.

Although our calculations suggest that the Coulomb interaction in
suspended graphene is not strong enough to open a dynamical gap, the
possibility of dynamical fermion gap generation can not be entirely
precluded. Indeed, there do exist several possible mechanisms for
the dynamical generation of fermion gap. For instance, the Dirac
fermions can acquire a tiny bare gap for some reasons, such as
Kekule distortion \cite{Chamon} and spin-orbit coupling
\cite{Kane05}. When this happens, a large dynamical fermion gap can
be generated by a relatively weak Coulomb interaction
\cite{Zhang11}, which then produces properties analogous to the
remarkable phenomena of QCD \cite{Zhang11}. Moreover, the additional
on-site repulsive interaction may help to generate a dynamical gap
even if the Coulomb interaction itself is not sufficiently strong
\cite{Liu09, Gamayun10, LiuWang}. Finally, it is well known that an
external magnetic field perpendicular to the graphene plane can lead
to dynamical gap generation and insulating phase even at
infinitesimal coupling \cite{Khveshchenko01B, Gorbar02}.

\ack{J.R.W. thanks J. Wang and W. Li for very helpful discussions.
G.Z.L. acknowledges the financial support by the National Natural
Science Foundation of China under grant No. 11074234 and the
Visitors Program of MPIPKS at Dresden.}

\end{document}